\documentclass[numbers]{cas-sc}
\usepackage[numbers,sort&compress]{natbib}

\def\tsc#1{\csdef{#1}{\textsc{\lowercase{#1}}\xspace}}
\tsc{WGM}
\tsc{QE}

\begin{document}
\let\WriteBookmarks\relax
\def\floatpagepagefraction{1}
\def\textpagefraction{.001}

\shorttitle{Flattened vGRF Waveforms in Gait of Women with Knee OA}

\shortauthors{G. Bouchouras et al.}

\title [mode = title]{Vertical Ground Reaction Forces Waveform Flattening during Gait in Women with Knee Osteoarthritis}  

\tnotemark[1] 


%

\author[1]{Georgios Bouchouras}

\cormark[1]


\ead{gbouchouras@mitropolitiko.edu.gr}

\ead[url]{https://giorgosbouh.github.io/github-portfolio/}


\affiliation[1]{organization={Faculty of Sport Sciences and Physical Education, Metropolitan College},
            addressline={14 Venizelou Str}, 
            city={Thessaloniki},
            postcode={54624}, 
            state={Central Macedonia},
            country={Greece}}

\author[1]{Georgios Sofianidis}


\ead{gsofianidis@mitropolitiko.edu.gr}



\affiliation[1]{organization={Metropolitan College},
  city={Thessaloniki}, country={Greece}}


\author[1]{Syragoula Charisi}


\ead{scharisi@mitropolitiko.edu.gr}



\author[2]{Charalampos Pavlopoulos}


\ead{babispaul@hotmail.com}


\affiliation[2]{organization={2nd Orthopaedic Department, Gennimatas Hospital, Aristotle University of Thessaloniki},
            addressline={41 Ethnikis Amynis Str}, 
            city={Thessaloniki},
            postcode={54635}, 
            state={Central Macedonia},
            country={Greece}}


\author[3]{Vassilia Hatzitaki}


\ead{vaso1@phed.auth.gr}


\affiliation[3]{organization={Laboratory of Motor Behavior and Adapted Physical Activity, Department of Physical Education and Sport Science, Aristotle University of Thessaloniki},
            city={Thermi},
            postcode={57001}, 
            state={Central Macedonia},
            country={Greece}}

\author[4]{Efthimios Samoladas}


\ead{msamolad@auth.gr}


\affiliation[4]{organization={1st Orthopaedic Clinic, Georgios Papanikolaou General Hospital, Aristotle University of Thessaloniki},
            addressline={Leoforos Papanikolaou}, 
            city={Thessaloniki},
            postcode={57010}, 
            state={Central Macedonia},
            country={Greece}}



\begin{abstract}
Background. Knee Osteoarthritis (OA) is a common chronic joint condition, and its prevalence increases with age. This study aims to examine whether flattened vertical ground reaction force (vGRF) waveforms and reduced knee range of motion (RoM) occur together during gait as compensatory strategies to maintain gait speed. Methods. Twelve women with knee OA and twelve healthy women of the same age completed the Western Ontario and McMaster University Index (WOMAC) to assess self-reported pain, stiffness, and function. The groups were divided into two groups: OA vs. control  2 limbs or left and right in the Control group. A mixed-design ANOVA was used to examine differences in vertical ground response forces (VGRFs) peaks, minimum VGRF, anterior-posterior weight acceptance (ADWA) and propulsive force (ADPO), knee RoM, and gait speeds. Results. In the OA group, the mean Peak 1 vGFR was 1.109 (SD = 0.05) for the right leg (p  0.05), while the mean min vGFFR was 0.87 (SD=0.04) for the left leg. The OA leg exhibited a mean ADWA of 0.23  0.04 kg/BW, which was significantly lower than the control group's right leg (0.28 0.09 kg/bw, p0.05). No group differences in gait velocity were detected. Conclusions. We interpret the flattening of the vFGFR waveform and the reduction in knee RoM as components of an adaptive, yet potentially maladaptive, motor strategy.\end{abstract}


\begin{highlights}
\item Women with knee OA show flattened vGRF, low push-off force, and less knee RoM.
\item Reduced knee RoM leads to compensatory gait that may increase joint stress.
\item Sustained loading without normal cycles may accelerate joint degeneration.
\end{highlights}


\maketitle

\section{Introduction}\label{}
Osteoarthritis (OA) is a common chronic joint condition. An estimated 20\% of all adults are affected by the disease, and its prevalence increases with age \citep{Badley1995} of the joints’ mechanical alignment, pain, functional dependency, and problems associated with gait disruption \citep{Baliunas2002}, as well as changes in ground reaction forces (GRF) \citep{Esrafilian2013}, are considered the most common complications of knee osteoarthritis (OA) \citep{Shafizadegan2016}. Gait disruption reduces self-sufficient ability, impedes work and family life, imposes concomitant physical and mental illnesses, and increases the risk of mortality \citep{Salzman2011} as it can lead to falls \citep{Rubenstein2006}.

Vertical ground reaction forces (vGRF) during gait are critical biomechanical indicators used to evaluate how forces are transmitted through the body during gait. The vGRF waveform consists of two peaks: the passive (weight acceptance) peak and the active (push off) peak. The passive peak (Peak 1) is the result of the impact of the foot with the ground, while the active peak (Peak 2) results from the active force applied by the foot to the ground as it drives away \citep{Jiang2020}. The magnitude of these peaks influences the loads experienced in the joints and muscles of the lower extremity and can result in the development or exacerbation of musculoskeletal overuse injuries and conditions such as osteoarthritis \citep{Shafizadegan2016}. These peaks reflect how the body absorbs impact and generates propulsion, respectively. Flattening of this waveform, where distinct peak amplitudes become less pronounced, indicates altered joint loading and may suggest compromised joint function or degeneration. \citep{Costello2021}.

Flattening of the vGRF waveform is commonly observed in individuals with osteoarthritis and is considered an indicator of joint dysfunction. In this context, flattening describes a reduction in the amplitude between the two characteristic vGRF peaks and the mid-stance minimum, resulting in a less pronounced, more level waveform. However, exploration of how this flattened waveform correlates with other key biomechanical factors, such as knee range of motion (RoM) and gait speed, remains limited. Knee RoM often decreases in OA due to joint stiffness and pain, which limits the joint's ability to absorb impact and generate force during gait \citep{AlZahrani2002}. This reduction in RoM may be exacerbated by increased muscle co-contraction, which not only increases joint loads, but also contributes to accelerated cartilage damage \citep{Bouchouras2015a, HubleyKozey2009}. This co-contraction may enhance joint stability, but could also contribute to faster structural damage and a higher likelihood of progressing to total knee arthroplasty \citep{Hodges2016,Schipplein1991}. However, the potential consequences of these compensatory mechanisms need to be carefully considered. Although these mechanisms may help individuals with knee OA perform essential daily activities, their long-term impact on joint health and the progression of the disease remains a critical concern \citep{Bouchouras2015a, HubleyKozey2009}. Thus, understanding these compensatory mechanisms is crucial for assessing their impact on joint health and the progression of osteoarthritis. It is well known that knee OA results in joint degeneration, which is characterized by cartilage loss and increased discomfort. For this baseline degeneration, the following compensatory mechanisms (flat waveform and reduction of knee RoM), if detected, could contribute and accelerate cartilage breakdown. This study aims to determine whether flat vGRF force waveforms and restricted knee range of motion appear simultaneously during gait, as compensatory mechanisms generated to maintain the same gait speed, both of which may contribute to joint degeneration in osteoarthritis patients.

The study hypothesizes that individuals with knee osteoarthritis (OA) will show a reduced amplitude of vGRF during gait, indicating flattening of the vGRF waveform, along with a smaller range of motion of the knee joint and slower gait speeds compared to healthy controls. These variables will be examined through both between-group comparisons (OA vs. control participants) and within-group comparisons (affected vs. non-affected limb in the OA group.

By investigating these aspects, this study seeks to offer a deeper understanding of how alterations in vGRF waveforms relate to other critical gait variables in individuals with knee OA, potentially highlighting novel areas for intervention and management.

\section{Research methodology}
Twelve women with knee OA (age 64.02 ± 4.13 years, height 160.01 ± 3.02 cm, and mass 76.20 ± 8.83 kg) and twelve healthy women of the same age (age 62.19 ± 4.02 years, height 161.83 ± 4.27 cm, and mass 76.20 ± 8.88 kg) volunteered to participate in this study. Women with knee OA showed unilateral grade II or III OA in the medial tibiofemoral compartment of the right knee joint, as evidenced by radiographic assessment using Kellgren and Lawrence criteria \citep{Kellgren1957}. Finally, none of them reported any pain or discomfort in the left knee and were diagnosed by an orthopedic surgeon with grade I or II unilateral osteoarthritis in the right knee. Participants of the control group had no musculoskeletal or neuromuscular pain or any injury of the knee or hip joint as assessed by an orthopedic surgeon. They were regularly active, but did not participate in systematic training programs. All women completed the Western Ontario and McMaster University Index (WOMAC) to assess self-reported pain, stiffness, and function \citep{Bellamy1988}. The study was conducted in accordance with the ethical principles outlined in the Declaration of Helsinki. All participants were fully informed about the purpose and procedures of the study and provided written informed consent prior to participation. Data collection was carried out at the Laboratory of Motor Behavior and Adapted Physical Activity (School of Sport Sciences and Physical Education, Aristotle University of Thessaloniki).
\subsection{Instrumentation}
Two adjacent force plates (Balance Plate 6501, Bertec Corp., Columbus, USA) with a sampling frequency of 100 Hz were used to record the ground reaction forces during the task. A 10-camera Vicon system with sampling 100 Hz was used to capture the kinematic variables of the participants. (Bonita 3, Nexus Vicon, Oxford, UK). The ground reaction forces and markers 3D position coordinates were synchronously sampled using the Vicon system that was connected to an A/D card (MX Giganet Oxford, UK). 
\subsection{Procedure and data collection}
In order to collect the kinematic data, 16 passive reflective markers were attached to specific anthropometric landmarks according to the model of Davis \citep{Davis1991} for estimating the joint angles from the markers’ 3D position coordinates using the lower leg Plug-in-Gait model. Participants, after many familiarization trials, walked along a 4-meter walkway with two force plates embedded in the ground at their self-selected speed without shoes. Approximately three gait trials were acquired from each participant. A custom algorithm developed in MATLAB (Mathworks Inc., version 7.7.0471) was used to derive depended variables of the study. For each participant, the third trial out of three recorded trials was selected for analysis. This decision was made based on the assumption that by the third trial the participants had reached their optimal comfort in performing the task. We chose not to average the results from all three trials because doing so may not reflect a true single natural gait trial. Finally, all participants conducted tests barefoot to eliminate the confounding effects of footwear on ground reaction forces and ensure consistent force plate contact.

\subsubsection{The dependent variables of this study were}

{a) Gait speed for each leg, as calculated by the Nexus Software system. Gait speed for each leg was calculated using data from the Nexus Software system, which provided information on stride length and stride time. Stride length refers to the distance covered by the leg during a full gait cycle (from initial contact of one foot to the next contact of the same foot), while stride time refers to the time it takes to complete this cycle.
The formula used for calculating gait speed was:

\begin{equation}
    \text{Gait Speed} = \frac{\text{Stride length}}{\text{Stride time}}
\end{equation}

Specifically, this calculation was performed individually for each leg during the gait trials. Stride length was measured using the positional data from the passive reflective markers attached to anatomical landmarks, as recorded by the Nexus motion capture system. Stride time was derived from the time between consecutive ground contacts of the same leg, as identified by the vertical ground reaction force (vGRF) data. The beginning of the stance phase was defined as the moment when vGRF first rose above zero, and the end of the stance phase was marked when the vGRF returned to zero. In fact, initial contact and toe-off were defined using a threshold of 10 N to account for baseline noise in the vGRF signal. Using this information, the stride time was precisely determined by measuring the interval between successive phases of stance of the same leg. The resulting gait speed was expressed in meters per second (m/s) for each leg and used as a dependent variable in the analysis. The calculations took into account any potential differences in leg length and gait pattern between the two groups (OA and control) and between the limbs (right and left) of each participant.
In this study, we chose not to standardize gait speeds between individuals with knee OA and those with normal knees, but instead allowed participants to walk at their self-selected comfortable speed. This approach ensures ecological validity, as it reflects natural gait behavior rather than an imposed constraint. Gait speed is an important biomechanical variable, and OA patients often adjust their pace as a compensatory mechanism to minimize discomfort. Standardizing speed could obscure these adaptations and lead to less meaningful interpretations \citep{Harkey2021}. Additionally, self-selected speed is commonly used in clinical assessments as an indicator of mobility and functional capacity, making our findings more relevant to real-world applications. By allowing participants to walk at their preferred pace, we capture the true impact of OA on gait rather than imposing an artificial condition that may not accurately represent their functional abilities.

b) vGRF's: 
The force plates recorded the vGRF signals as participants walked at their self-selected speed. These signals were sampled at 100 Hz, which means that data points were collected 100 times per second, providing high-resolution force data throughout the stance phase of each gait cycle. The vGRF data was low-pass filtered using a Butterworth digital filter (fourth-order, dual pass, cutoff at 6 Hz) to remove high-frequency noise and ensure smooth and interpretable force waveforms. The filtered vGRF data were then normalized to each participant's body weight (expressed as a ratio to body weight) to account for differences in body mass between participants.}

b1. First Peak vGRF (Peak 1 vGRF): The first peak of the vGRF  occurred during the weight acceptance phase, shortly after initial contact of the foot with the ground. This peak represents the passive response as the body absorbs the force of landing. To calculate this value, the vGRF waveform was visually and computationally inspected, and the maximum value during the early stance phase (just after initial contact) was identified as the first peak (Fig 1). 

b.2 Second Peak vGRF (Peak 2 vGRF): The second peak of the vGRF (Peak 2) occurred during the push-off phase, when the foot actively pushes off the ground to propel the body forward. The second peak was identified as the maximum vGRF value that occurred near the end of the stance phase, just before the foot leaves the ground. This peak results from the active force generated by the foot pushing off the ground as the body prepares for the next step. This peak reflects the propulsion phase of gait (Fig 1).
b3. Minimum vGRF (min vGRF): The minimum vGRF was the lowest value recorded between the first peak and the second peak during the stance phase of gait. This value represents the point of minimum loading during the stance phase of the gait cycle, reflecting the transition between the passive weight acceptance phase and the active push-off phase. To determine this value, the vGRF waveform was inspected between the first and second peaks, and the lowest point was selected as the minimum. This value helps quantify the unloading phase, where the forces acting on the joint are reduced before the foot pushes off the ground (Fig 1).

c) The amplitude difference for both the weight acceptance (ADWA) and push-off phases (ADPO) was calculated as the difference between the respective peak (first or second peak) and the min vGRF that occurred between the two peaks. This value captures how much the vGRF changes from the minimum loading to the peak loading during each phase of gait (Fig 1).

Overall Procedure:
The first and second peaks were calculated for each test for both the right and left legs. These peaks were critical for understanding the loading and unloading patterns during gait in both the OA and control groups. To ensure that differences in body weight did not affect the analysis, the vGRF values were normalized to each participant’s body weight (BW). The normalized vGRF was expressed as a ratio (Newtons/BW), which allows for direct comparisons between individuals regardless of their mass. Additionally, to enhance the reader's understanding of the force, the vGRF values were changed from Newtons to kilograms during the analysis.

d) Peak-to-peak amplitude of the knee angular displacement (deg) in the sagittal plane for stance phase. Knee range of motion (RoM) during the stance phase was calculated as the peak-to-peak amplitude of angular displacement in the sagittal plane. This refers to the difference between the maximum and minimum knee angles observed during the stance phase. The maximum knee flexion angle typically occurs during the mid-stance phase, when the knee is most bent as the body passes over the foot. The minimum knee flexion angle (or maximum knee extension) typically occurs during the initial contact and push-off phases, when the knee is more extended. Knee RoM was calculated by subtracting the minimum knee angle (extension) from the maximum knee angle (flexion), providing the total angular displacement (in degrees) of the knee during the stance phase. Knee RoM was analyzed for both the right and left legs in all participants. For each participant, the RoM was calculated from their recorded kinematic data for three trials, with the final trial selected for analysis. The calculated knee RoM values were then used to compare within-subject differences (between the affected and non-affected leg in the OA group) and between-group differences (between the OA and control groups).
\subsection{Design and statistical analysis}

A mixed-design ANOVA (2 groups: OA vs. control × 2 limbs) was used to examine differences in vertical ground reaction force (vGRF) peaks, minimum vGRF, anterior-posterior weight acceptance (ADWA), anterior-posterior propulsive force (ADPO), knee range of motion (RoM), and gait speed. This analysis, implemented using the repeated measures function in SPSS, assessed both between-group (OA vs. control) and within-group (affected vs. non-affected limb in the OA group or left and right in the Control group) effects. 
Comparisons between groups (OA versus control) were performed using the factor between subjects of mixed-design ANOVA, with both legs included as repeated measures. 
The effects within the group (legs) were evaluated using the factor within the subjects in the mixed design ANOVA to determine whether the presence of knee osteoarthritis resulted in unilateral impairments, as evidenced by asymmetrical biomechanical patterns.

The level of statistical significance was set at p < 0.05. Partial eta-squared (h²p) was calculated to estimate effect sizes, following Cohen’s guidelines \citep{Cohen1988} 0.2 for a small effect, 0.5 for a moderate effect, and 0.8 for a large effect. Post-hoc tests were conducted to further analyze any significant interactions or main effects.

All parametric assumptions were met, as confirmed through Shapiro-Wilk and Levene’s tests for normality and homogeneity of variance. Data from three participants in the OA group were excluded due to incomplete trials, leaving a final sample of nine participants in the OA group and twelve in the control group.

All statistical analyses were conducted using SPSS 23 (Armonk, NY: IBM Corp.). Results are reported with corresponding mean values, standard deviations, p-values, and eta partial squared (h²p) where applicable.

\begin{figure}[h!]
    \centering
    \includegraphics[width=\textwidth]{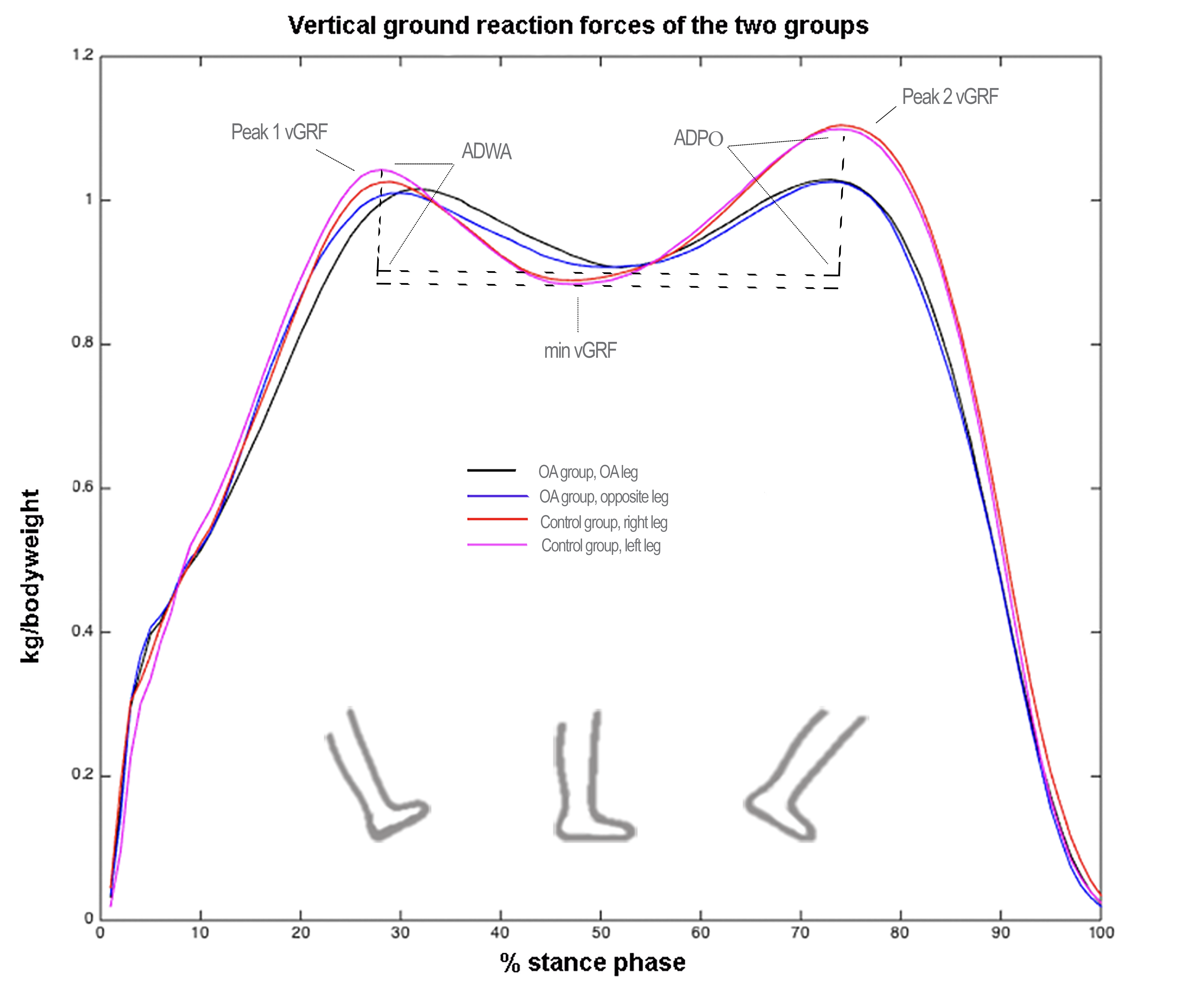} 
    \caption{Vertical ground reaction forces of each leg for the two groups (mean values of all participants)}
    \label{fig:vgrf}
\end{figure}

\section{Results}

\subsection{Peak 1 vGRF}

Within-Subjects: The repeated measures ANOVA for within-subject effects showed a significant difference between the two legs (F(1, 19) = 5.583, p = 0.029, h²p = 0.227). However, the interaction between "Leg" and "Group" (OA vs. Control) was non-significant (F(1, 19) = 0.477, p = 0.498, h²p = 0.024), indicating consistent effects across legs (table 2).

Between-Subject analysis revealed no significant difference in the Peak 1 vGRF between the OA and Control groups (F(1, 19) = 2.00, p = 0.17, h²p = 0.09). Descriptive statistics showed that in the Control group, mean Peak 1 vGRF was 1.13 (SD = 0.05) for the right leg and 1.16 (SD = 0.06) for the left leg. In the OA group, the mean Peak 1 vGRF was 1.109 (SD = 0.05) for the right leg and 1.12 (SD = 0.06) for the left leg (Table 1).

\subsection{Peak 2 vGRF}

Within-Subjects: The repeated measures ANOVA showed no significant difference between the right and left legs in the second Peak 2 vGRF (F(1, 19) = 0.77, p = 0.39, h²p = 0.03). The interaction between "Leg" and "Group" was also non-significant (F(1, 19) = 0.19, p = 0.66, h²p = 0.01, table 2).

Between-Subjects: The between-subjects analysis revealed a significant effect of group (OA vs. Control) on the
Peak 2 vGRF (F(1, 19) = 7.76, p = 0.01, h²p = 0.29). The descriptive statistics showed that in the control group, the
mean second peak of vGRF was 1.12 (SD = 0.06) for the right leg and 1.12 (SD = 0.06) for the left leg. In the OA
group, the mean Peak 2 vGRF was 1.05 (SD = 0.06) for the OA leg and 1.04 (SD = 0.06) for the opposite leg. Post hoc comparisons revealed that the right leg of the control group
had a significantly higher vGRF than the opposite leg of the OA group(p=0.04). Lastly, the comparison between the
opposite leg of the OA group and the left leg of the Control group approached significance (p = 0.05), indicating a
trend toward a higher vGRF in the Control group (Fig. 2, Table 1).

\begin{figure}[h!]
    \centering
    \begin{minipage}{0.49\textwidth}
        \centering
        \includegraphics[width=\textwidth]{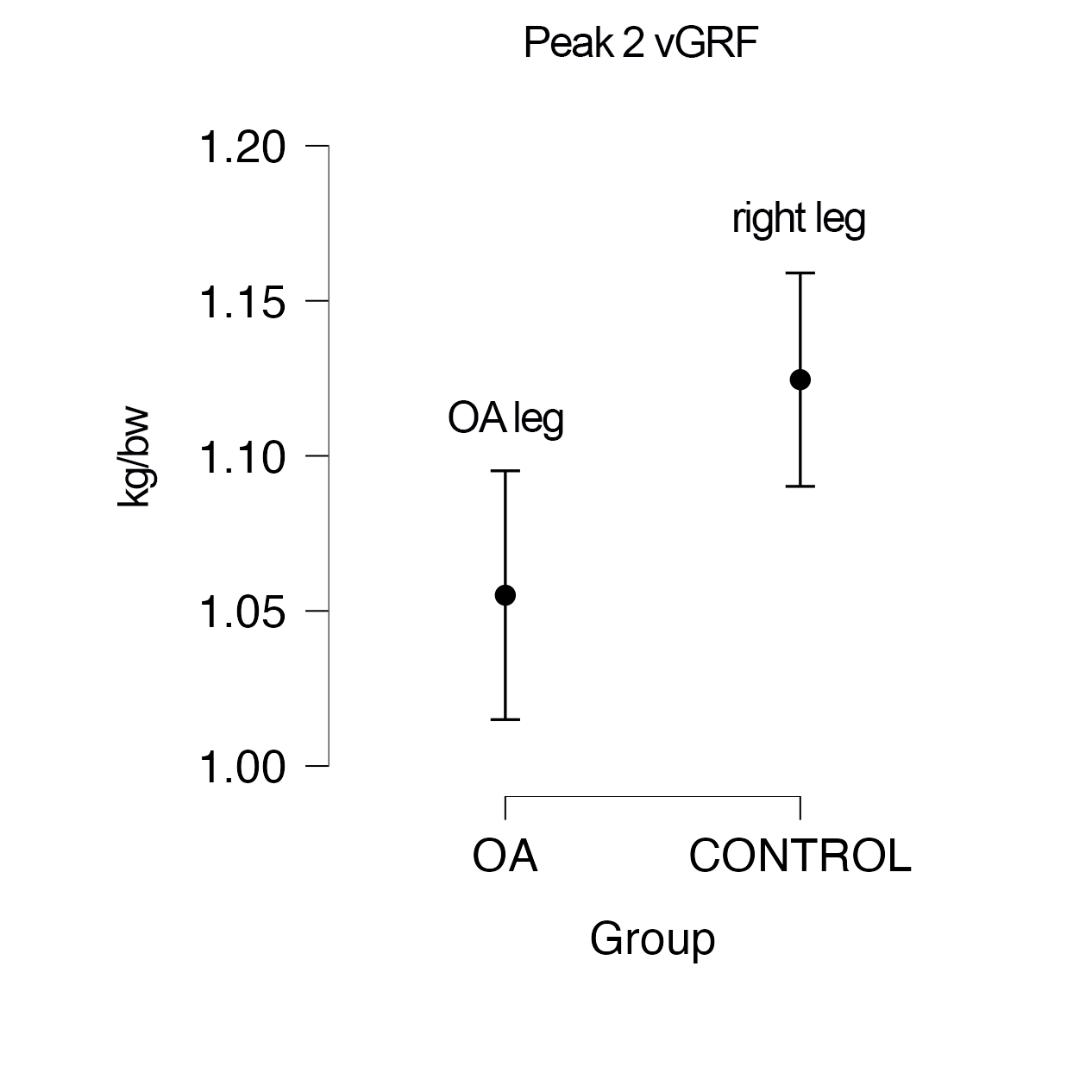} 
       
    \end{minipage}
    \hfill
    \begin{minipage}{0.49\textwidth}
        \centering
        \includegraphics[width=\textwidth]{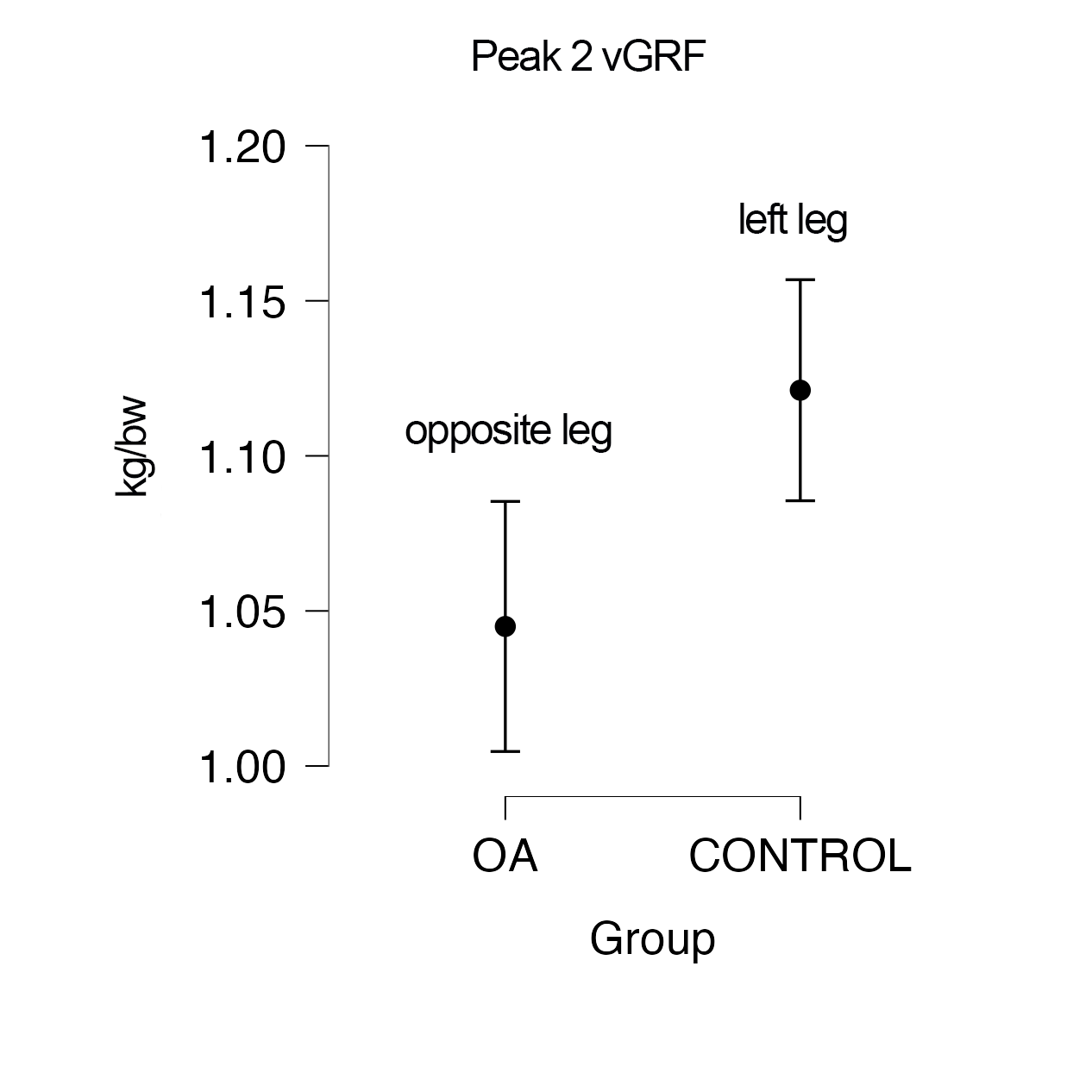} 
      
    \end{minipage}
    \caption{Second peak value of vertical ground reaction for both legs in between groups comparison.}
    \label{fig:second_peak_vgrf}
\end{figure}

\subsection{Minimun vGRF}

Within-Subjects: The repeated measures ANOVA revealed no significant differences between legs (F(1, 19) = 0.20, p = 0.65, h²p = 0.01). The interaction between 'Leg' and 'Group' was also nonsignificant (F (1, 19) = 0.00, p = 0.95, h²p = 0.00, table 2).

Between-Subjects: There were no significant differences in min vGRF between the OA and control groups (F (1, 19) = 0.98, p = 0.33, h²p = 0.04). The descriptive statistics showed that the control group had a mean min vGRF of 0.85 (SD = 0.06) for the right leg and 0.84 (SD = 0.06) for the left leg. The OA group had a mean min vGRF of 0.87 (SD = 0.05) for the right leg and 0.87 (SD = 0.04) for the left leg (Table 1).

\subsection{ADWA}

Within-Subjects: The repeated measures ANOVA for within-subject effects revealed a significant difference between the right and left legs (F(1, 19) = 77.88, p < 0.00, h²p = 0.80), indicating that the amplitude was significantly higher in the right leg compared to the left leg across participants. However, the interaction between Leg and Group (OA versus Control) was not significant (F (1, 19) = 1.86, p = 0.18, h²p = 0.08), suggesting that the effect of leg on weight acceptance amplitude was consistent between the OA and Control groups (table 2).

Between-Subjects: For between-subjects effects, the analysis revealed no significant difference between the OA and Control groups (F(1, 19) = 1.95, p = 0.17, h²p = 0.09). Descriptive statistics showed that, for the right leg, the mean ADWA was 0.28 (SD = 0.09) in the Control group and 0.23 (SD = 0.04) in the OA group. For the left leg, the mean ADWA was 0.11 (SD = 0.04) in the Control group and 0.11 (SD = 0.03) in the OA group. These results suggest that there are no substantial differences between groups in the weight acceptance amplitude (table 1). 

However, post hoc comparisons revealed several significant differences. In the OA group, the OA leg exhibited an ADWA of 0.23 ± 0.04 kg/BW, which was significantly lower than the control group’s right leg (0.28 ± 0.09 kg/BW, p < 0.05). Additionally, within the OA group, the opposite leg showed a significantly lower ADWA (0.11 ± 0.03 kg/BW) compared to the OA leg (p < 0.05), suggesting asymmetrical work absorption likely due to compensatory mechanisms. In the Control group, although a substantial difference in the means was observed, there was no significant difference in ADWA between the right leg (0.28 ± 0.09 kg/BW) and the left leg (0.11 ± 0.04 kg/BW), indicating more balanced work absorption compared to the OA group.

\begin{figure}[h!]
    \centering
    \begin{minipage}{0.49\textwidth}
        \centering
        \includegraphics[width=\textwidth]{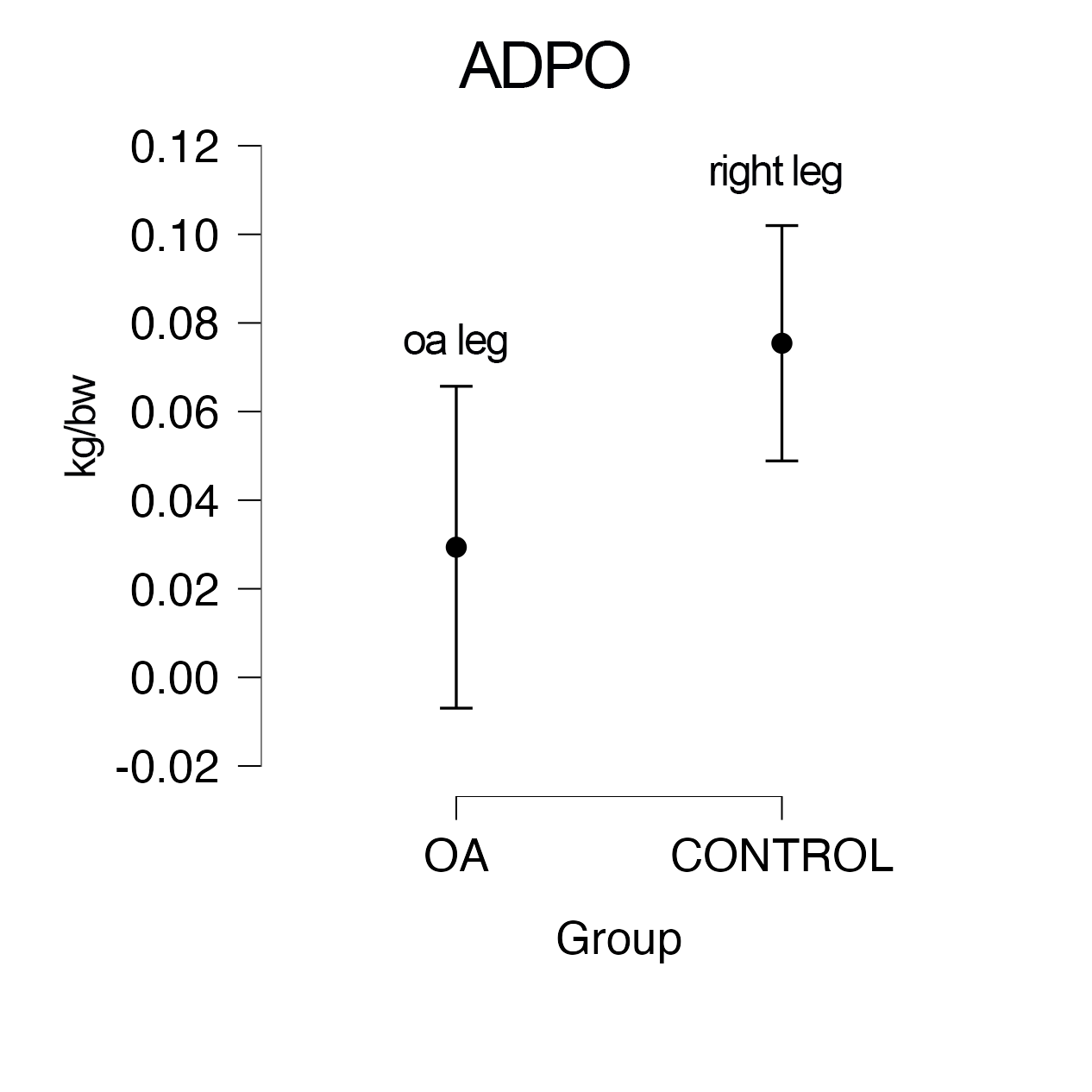} 
        
    \end{minipage}
    \hfill
    \begin{minipage}{0.49\textwidth}
        \centering
        \includegraphics[width=\textwidth]{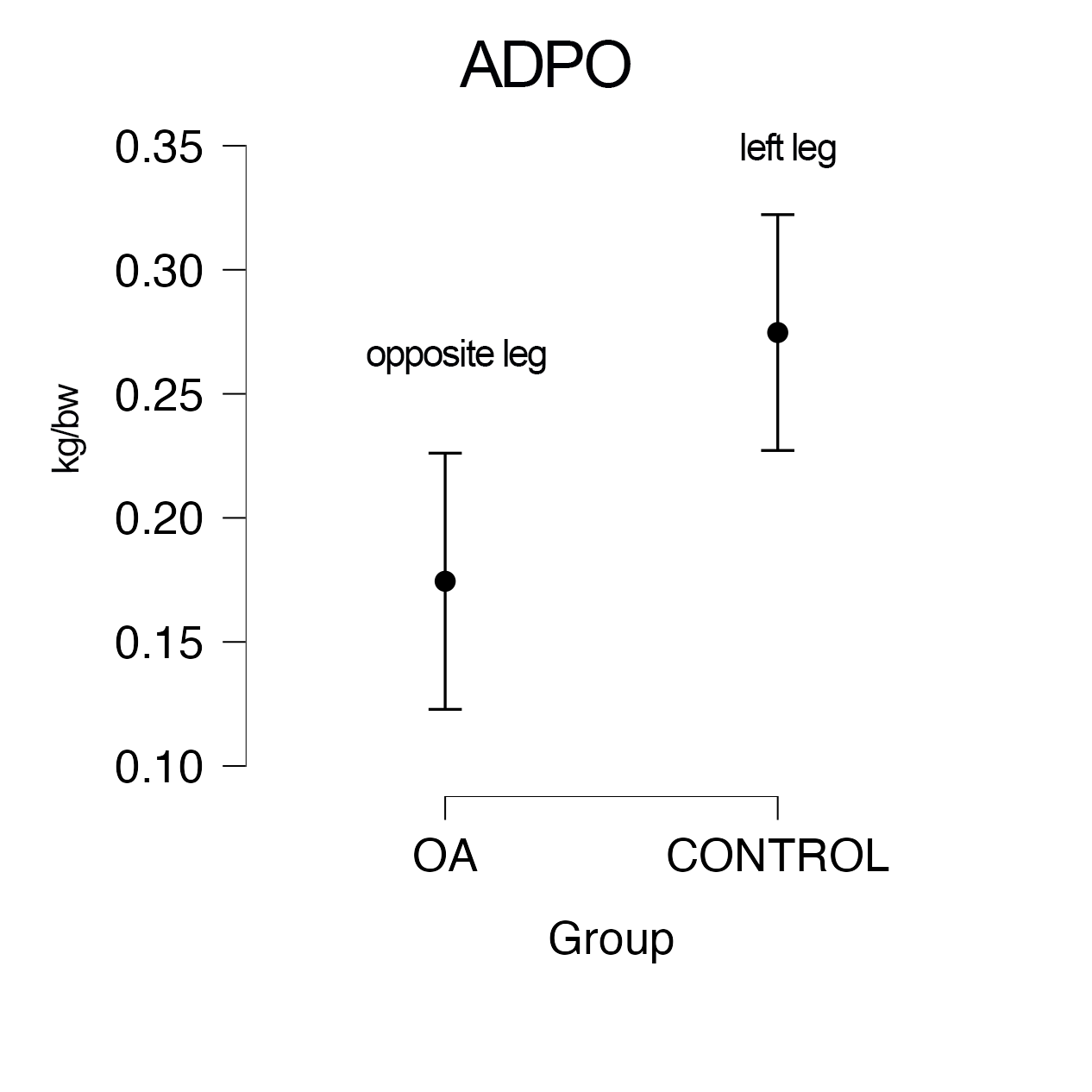} 
       
    \end{minipage}
    \caption{Amplitude difference of the vGRF during push-off phase for both legs legs in between groups comparison.}
    \label{fig:adpo_vgrf_push_off}
\end{figure}

\subsection{ADPO}

Within-Subjects: The repeated measures ANOVA for within-subject effects revealed a significant difference in push-off amplitude between the right and left legs (F(1, 19) = 92.35, p < 0.00, h²p = 0.82), indicating that the amplitude was significantly higher in the left leg compared to the right leg. However, the interaction between Leg and Groups (OA vs Control) was not significant (F (1, 19) = 2.28, p = 0.14, h²p = 0.10), suggesting that the effect of leg on push-off amplitude was consistent between the OA and Control groups (table 2).

Between-Subjects: The between-subjects analysis revealed a significant effect of condition on ADPO (F(1, 19) = 9.18, p = 0.00, h²p = 0.32), indicating that the OA group had significantly lower push-off amplitude compared to the Control group. The descriptive statistics showed that, for the right leg, the mean ADPO was 0.07 (SD = 0.047) in the control group and 0.029 (SD = 0.056) in the OA group. For the left leg, the mean ADPO was 0.27 (SD = 0.08) in the Control group and 0.17 (SD = 0.07) in the OA group. These results suggest a substantial difference in push-off amplitude between the OA and Control groups (table 1).

Furthermore, post hoc analysis revealed a number of notable variations. The opposite leg in the OA group had significantly higher push-off amplitude than the OA leg (p<0.00). The left control leg had significantly higher push-off amplitude than the opposite leg (p<0.00) and the OA leg of the OA group (p<0.00). Furthermore, the control right leg had a higher pushoff amplitude compared to the opposite leg of the OA (p = 0.01). Lastly, the Control left leg had a higher amplitude than the Control right leg (p<0.00, Fig 3).

\subsection{Knee RoM}

Within-Subjects: Repeated measures ANOVA did not reveal significant differences between the right and left knees in terms of RoM (F(1, 14) = 0.454, p = 0.511, h²p = 0.031). The interaction between 'Leg' and 'Group' was also nonsignificant (F (1, 14) = 1.611, p = 0.225, h²p = 0.103).

Between-Subjects: There was a significant group effect on knee RoM (F(1, 14) = 9.84, p = 0.00, h²p = 0.41). Descriptive statistics showed that in the Control group, the mean RoM was 51.79 degrees (SD = 8.31) for the right knee and 48.68 degrees (SD = 20.73) for the left knee. In the OA group, the mean RoM was 27.04 degrees (SD = 11.93) for the OA knee and 37.18 degrees (SD = 18.10) for the opposite knee (Fig 4, table 1). 

\begin{figure}[h!]
    \centering
    \begin{minipage}{0.49\textwidth}
        \centering
        \includegraphics[width=\textwidth]{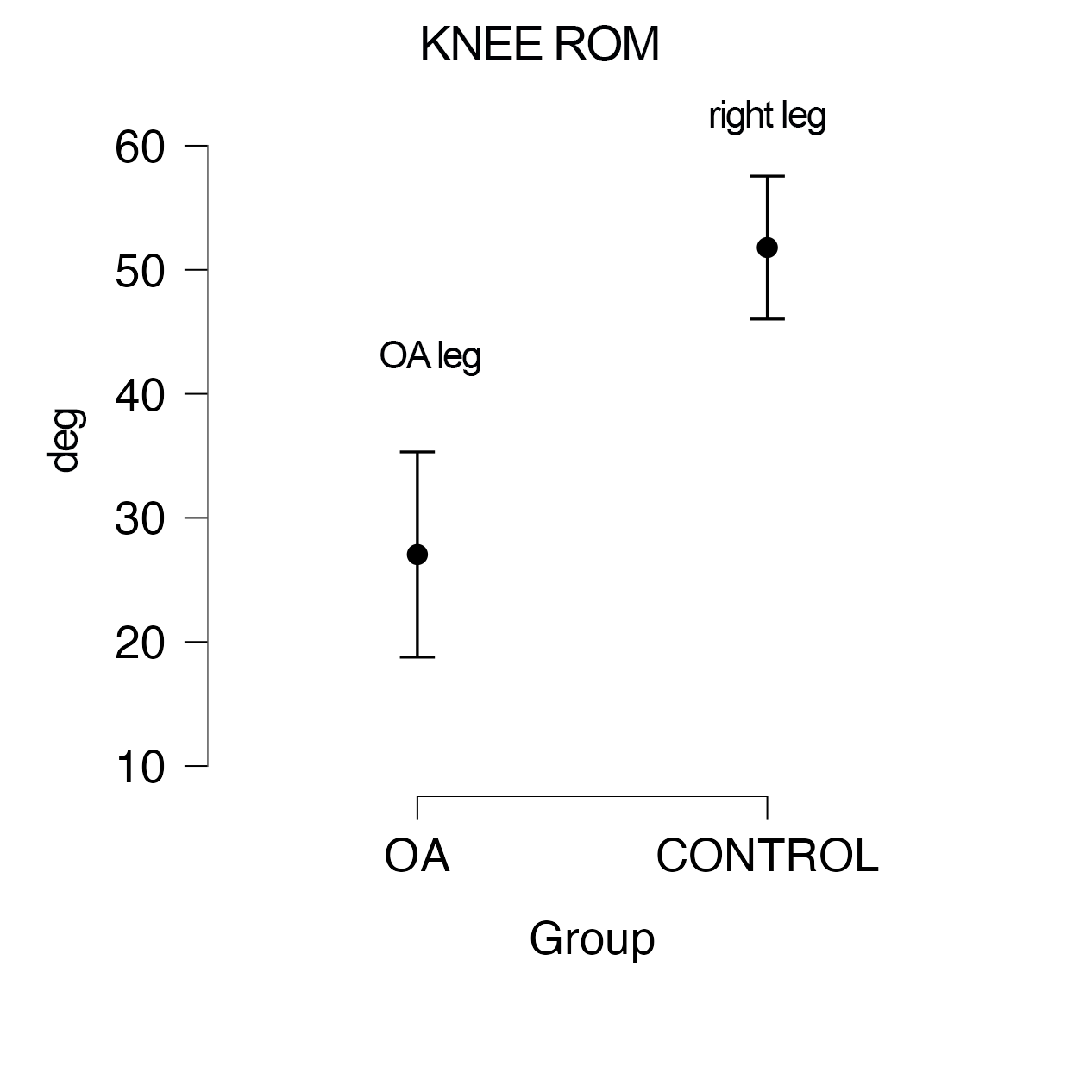} 
       
    \end{minipage}
    \hfill
    \begin{minipage}{0.49\textwidth}
        \centering
        \includegraphics[width=\textwidth]{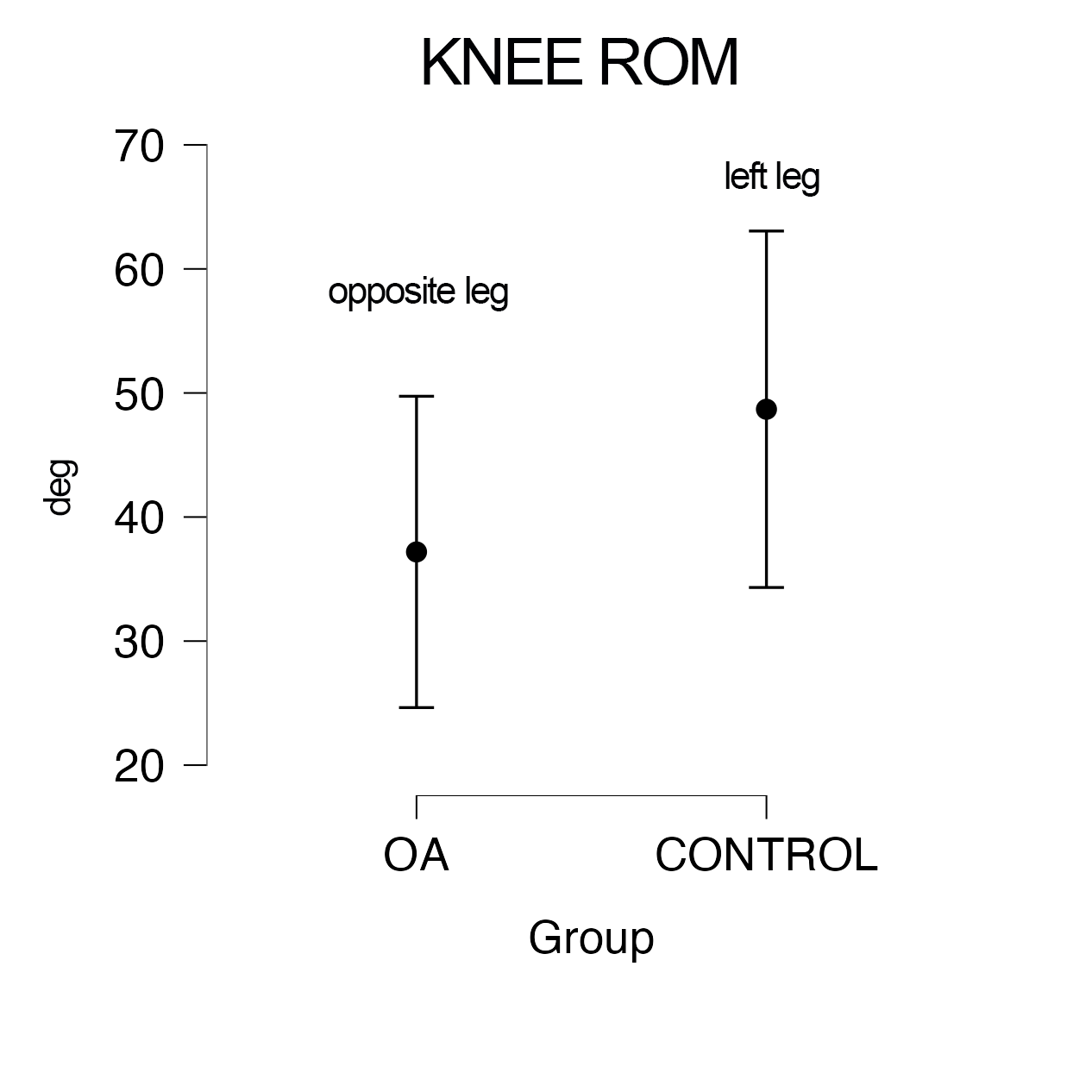} 
       
    \end{minipage}
    \caption{Knee range of motion in the sagittal plane for both legs in between groups comparison.}
    \label{fig:knee_rom}
\end{figure}

\subsection{Gait speed}
No significant differences in gait speed were observed between or within the groups. The OA leg (0.89 ± 0.06 m/s) and the opposite leg (0.89 ± 0.10 m/s) showed comparable values (p = 0.40), indicating symmetric gait speed within the OA group. Similarly, the control group did not show significant side-to-side differences (Control Right: 0.90 ± 0.10 m/s; Control Left: 0.93 ± 0.11 m/s; p = 0.40). Comparisons between groups also did not reveal statistically significant differences in gait speed (OA leg versus control right: p = 0.75; Opposite leg versus control left: p = 0.40).

\begin{table}[h]
\centering
\caption{Between-group leg-specific comparison (OA Leg vs Control Right Leg; Opposite Leg vs Control Left Leg)}
\begin{tabular}{lcccccc}
\toprule
Variable & OA Leg & Opp. Leg & Ctrl Right & Ctrl Left & p$_1$ & p$_2$ \\
\midrule
Peak 1 vGRF (kg/BW) & 1.10 ± 0.05 & 1.12 ± 0.06 & 1.13 ± 0.05 & 1.16 ± 0.06 & 0.30 & 0.21 \\
Peak 2 vGRF (kg/BW) & 1.05 ± 0.06 & 1.04 ± 0.06 & 1.12 ± 0.06 & 1.12 ± 0.06 & \textbf{0.04}$^*$ & \textbf{0.05}$^*$ \\
Min vGRF (kg/BW) & 1.02 ± 0.05 & 1.01 ± 0.06 & 1.04 ± 0.05 & 1.05 ± 0.05 & 0.41 & 0.33 \\
ADWA (kg/BW) & 0.23 ± 0.04 & 0.11 ± 0.03 & 0.28 ± 0.09 & 0.11 ± 0.04 & \textbf{0.05}$^*$ & 0.87 \\
ADPO (kg/BW) & 0.02 ± 0.05 & 0.17 ± 0.07 & 0.07 ± 0.04 & 0.27 ± 0.08 & \textbf{0.01}$^*$ & \textbf{0.00}$^*$ \\
Knee RoM (deg) & 27.04 ± 11.93 & 37.18 ± 18.10 & 51.79 ± 8.31 & 48.68 ± 20.73 & \textbf{0.00}$^*$ & \textbf{0.01}$^*$ \\
Gait Speed (m/sec) & 0.89 ± 0.06 & 0.89 ± 0.10 & 0.90 ± 0.10 & 0.93 ± 0.11 & 0.75 & 0.40 \\
\bottomrule
\end{tabular}
\vspace{0.5em}

\small{p$_1$: OA leg vs. Control Right Leg, p$_2$: Opposite leg vs. Control Left Leg. $^*$Significant difference (p < 0.05).}
\end{table}

\begin{table}[h]
\centering
\caption{Within-group comparison of biomechanical variables (OA Leg vs Opposite Leg and Control Right vs Left Leg)}
\begin{tabular}{lcccccc}
\toprule
Variable & OA Leg & Opposite Leg & Control R & Control L & p$_{OA}$ & p$_{Control}$ \\
\midrule
Peak 1 vGRF (kg/BW) & 1.10 ± 0.05 & 1.12 ± 0.06 & 1.13 ± 0.05 & 1.16 ± 0.06 & \textbf{0.03}$^\dagger$ & 0.20 \\
Peak 2 vGRF (kg/BW) & 1.05 ± 0.06 & 1.04 ± 0.06 & 1.12 ± 0.06 & 1.12 ± 0.06 & 0.39 & 0.66 \\
Min vGRF (kg/BW) & 1.02 ± 0.05 & 1.01 ± 0.06 & 1.04 ± 0.05 & 1.05 ± 0.05 & 0.65 & 0.33 \\
ADWA (kg/BW) & 0.23 ± 0.04 & 0.11 ± 0.03 & 0.28 ± 0.09 & 0.11 ± 0.04 & \textbf{0.02}$^\dagger$ & 0.18 \\
ADPO (kg/BW) & 0.02 ± 0.05 & 0.17 ± 0.07 & 0.07 ± 0.04 & 0.27 ± 0.08 & \textbf{<0.001}$^\dagger$ & \textbf{<0.001}$^\dagger$ \\
Knee RoM (deg) & 27.04 ± 11.93 & 37.18 ± 18.10 & 51.79 ± 8.31 & 48.68 ± 20.73 & 0.23 & 0.51 \\
Gait Speed (m/sec) & 0.89 ± 0.06 & 0.89 ± 0.10 & 0.90 ± 0.10 & 0.93 ± 0.11 & 0.40 & 0.40 \\
\bottomrule
\end{tabular}
\vspace{0.5em}

\small{p$_{OA}$: OA Leg vs Opposite Leg. \quad p$_{Control}$: Control Right vs Left Leg. \quad $^\dagger$p < 0.05.}
\end{table}

\section{Discussion}

In the present study, we examined whether women with knee osteoarthritis (OA) exhibit differences in vertical ground reaction force (vGRF) patterns, knee range of motion (RoM), and gait speed compared to healthy controls. We also explored asymmetries between the affected and unaffected limbs within the OA group. Our primary hypothesis regarding vGRF amplitude was confirmed: the OA group exhibited significantly reduced push-off forces (Peak 2 vGRF) and amplitude differences (ADPO) compared to controls. These results, along with diminished ADWA, support the presence of a flatter vGRF waveform and altered load distribution in individuals with knee OA.

This flattening, reflected in the reduced amplitude between the two characteristic vGRF peaks, is likely part of a broader compensatory loading strategy aimed at minimizing pain and joint stress. Although we did not measure stride-to-stride variability explicitly, the consistently diminished second peak across participants may suggest a more uniform or sustained loading profile. Articular cartilage relies on dynamic mechanical input to maintain its metabolic and structural health \citep{Griffin2005, Lafortune1996}, and deviations from cyclical loading—whether due to underloading, overloading, or prolonged mid-stance—may disrupt cartilage homeostasis. While our findings do not permit direct conclusions regarding joint overload or under-stimulation, the altered force profile observed in the OA group highlights the need for further investigation using inverse dynamics and kinetic modeling.

Our results align with prior studies reporting lower vGRF amplitudes and flattened waveforms in individuals with knee OA \citep{AlZahrani2002, Kiss2011, Zahraee2014}. These findings are clinically relevant, as sustained or asymmetric loading during gait has been linked to joint deterioration. For example, Costello et al. (2021) found prolonged mid-stance loading and flattened GRF curves in limbs with knee OA, which may reflect pain-avoidance strategies but can result in adverse biomechanical consequences over time.

Additionally, our study revealed that despite reduced push-off force and RoM, minimum vGRF values remained symmetric between the OA and contralateral legs, suggesting an effort to preserve bilateral unloading symmetry. This may be a conscious or subconscious strategy to maintain gait stability and balance. Interestingly, no group differences in gait speed were detected, indicating that women with knee OA were able to maintain functional gait pace through compensatory patterns—albeit at the cost of biomechanical efficiency.

The significantly lower knee RoM in the OA group further supports the notion of a pain-avoidance strategy. Reduced RoM likely limits shock absorption and propulsive force generation, reinforcing the observed vGRF flattening. Prior research has shown that joint stiffness, often accompanied by increased co-contraction, can contribute to elevated joint loads and accelerated degeneration \citep{HubleyKozey2009}.

Our findings are consistent with a growing body of evidence on compensatory mechanisms in knee OA. Tanpure et al. (2023) reported that even after total knee arthroplasty, patients demonstrated significant alterations in the contralateral limb’s kinematics, suggesting redistribution of load and potential overuse \citep{Tanpure2023}. Downie et al. (2025) further showed that during a simulated fall, individuals with knee OA exhibited increased knee flexion but reduced trunk and hip flexion—postural adaptations that may protect the joint yet mask underlying instability \citep{Downie2025}. Collectively, these studies illustrate the multifaceted nature of compensation in OA, encompassing both gait and reactive balance strategies.

We interpret the flattening of the vGRF waveform and the reduction in knee range of motion as components of an adaptive, yet potentially maladaptive, motor strategy. While such compensations may support safer, pain-reduced gait in the short term, they could limit the joint’s exposure to the mechanical stimuli essential for maintaining cartilage health and long-term function. Importantly, this variability in motor patterning should not be viewed solely as pathological, but as an expression of the body’s attempt to optimize movement under biomechanical constraint \citep{Bouchouras2015a, Hatzitaki2007, Hong2006, Edelman2001}. Addressing these altered loading patterns through targeted clinical strategies—such as gait retraining to improve joint mechanics \citep{goff2022physical}, orthotic support to redistribute forces, neuromuscular training to enhance motor control, and weight management to reduce joint stress—offers a pathway to preserve function and minimize joint degeneration. When integrated with individualized physical therapy, these interventions have the potential to mitigate the negative consequences of compensatory gait mechanics, improve quality of life, and slow the progression of osteoarthritis \citep{swami2024literature, goff2022physical}.

Despite the strength of the observed effects, several limitations must be acknowledged. First, the relatively small sample size is a common constraint in clinical gait research, particularly when studying individuals with specific conditions such as radiographically confirmed unilateral knee osteoarthritis (OA). Recruitment challenges due to strict inclusion criteria, as well as the study’s focus on women only, further limited the eligible population. Nonetheless, the observed effect sizes for key variables—Peak 2 vGRF (h²p = 0.29), ADPO (h²p = 0.32), and knee RoM (h²p = 0.36)—were moderate to large, suggesting that the study was sufficiently powered to detect both between- and within-group differences. Another limitation involves the scope of biomechanical analysis. The study focused solely on vertical ground reaction forces (vGRF), omitting mediolateral and anteroposterior components of force, which also contribute to comprehensive joint loading profiles. Additionally, electromyographic (EMG) data were not collected. While EMG could have provided insights into neuromuscular control and compensatory muscle activation patterns, its inclusion was beyond the scope of this study. The decision to focus on vGRF and knee kinematics was based on their direct relevance to joint compression forces and the progression of knee OA, as supported by prior research (Lafortune et al., 1996). It is also important to note that surface EMG introduces methodological complexities, including sensitivity to electrode placement, signal noise, and inter-subject variability due to skin impedance. Nevertheless, the absence of EMG data is acknowledged as a limitation, particularly given the known role of co-contraction in joint stability and compensatory strategies in OA gait. Future research should aim to incorporate both EMG and multidirectional ground reaction forces, along with inverse dynamics modeling, to provide a more comprehensive understanding of the neuromechanical adaptations associated with OA. Moreover, this study employed discrete time-point measures (e.g., Peaks 1 and 2, and minimum vGRF), which, while clinically meaningful, may not capture the full temporal complexity of gait mechanics. Future studies may benefit from using continuous statistical methods such as Statistical Parametric Mapping (SPM) to analyze full waveform data.

To our knowledge, this is one of the few studies that examines the precise vGRF waveform, knee range of motion (RoM), and gait speed in women with knee OA. It is crucial to identify whether these parameters appear simultaneously during gait, as their concurrent presence may significantly contribute to further joint degeneration. We observed that women with knee OA exhibit waveform flattening, diminished ability to generate propulsive force, and reduced knee range of motion during gait. Nevertheless, these alterations do not compromise overall task duration, thus allowing the execution of this daily activity with safety. It seems that fear of pain and the associated reduction in range of motion are factors contributing to the formation of a distinct pattern of gait execution in patients with OA. This pattern, which appears to be compensatory, can support efficient and safe movement, but at the same time may place additional strain on the joint. This, in turn, could disrupt the normal load-unload cycles essential for joint health.  Prolonged joint loading without healthy periodicity may lead to further degeneration. This compensatory strategy may represent one of the diverse motor patterns used by individuals with knee OA to accomplish daily activities.  Additionally, from a clinical perspective, it would be of interest to investigate whether a rehabilitation program based on these findings could help slow the progression of cartilage damage.




\end{document}